\documentclass[12pt]{article}
\usepackage{euscript,amsmath,amssymb,amsthm,epsfig,latexsym}

\newcommand{\be}{\begin{equation}}
\newcommand{\ee}{\end{equation}}
\newcommand{\bea}{\begin{eqnarray}}
\newcommand{\eea}{\end{eqnarray}}

\renewcommand{\phi}{\varphi}

\begin{document}
\title{\ Exact solutions of Friedmann equation}
\vspace{1cm}
\author{
E. A. Kuryanovich\\
\\
\small{\it Research and Educational Center} \\
\small{\it Steklov Mathematical Institute Russian Academy of Sciences, Moscow}\\
\small{\it email:\texttt{kurianovich@mail.ru}}}

\date {~}
\maketitle

\begin{center}
    \textbf{Abstract}
\end{center}

The cosmological Friedmann equation for the universe filled with a scalar field is reduced to a system of two equations of the first order, one of which is an equation with separable variables. For the second equation the exact solutions are given in closed form for potentials as constants and exponents. For the same equation exact solutions for quadratic potential are written in the form of a series in the spiral and attractor areas. Also exact solutions for very arbitrary potentials are given  in the neighborhood of endpoint and infinity. The existence of all these classical solutions is proven.

\newpage

\section{Introduction}

 Study of Friedmann equation for the gravitational field interacting with a scalar field plays an important role in modern cosmology. Within
inflationary theory an approximate analysis of Friedmann equations for free massive scalar field was conducted, see. eg [1, p. 236]. In the same paper, the phase trajectory of Friedmann equation with quadratic potential  was built [1, p. 237, fig. 5.3]. This trajectory is a spiral coming out of infinity and endlessly twists to the origin. In this work it has also been shown that the trajectory has a so-called "attractor" line, along which this trajectory moves under certain initial data for a long time. Thus, the phase trajectory for the Friedmann equation with quadratic potential can be divided into three areas: a neighborhood of infinity, spiral neighborhood of the origin and the "attractor" area, which is not always the case. In ref [2-4] various properties of Friedmann equation solutions have been investigated, and solutions for the potentials in the form of constant and exponent have been found. In ref [5] the solution of Friedmann equation with quadratic potential was built in the neighborhood of infinity in the form of generalized Dirichlet series.

This paper gives a detailed mathematical analysis of the Friedmann equations
 for a scalar field with quadratic potential in a spiral and
an attractor  areas. Also this equation is studied with the continuous slow
 growing arbitrary potential in the neighborhood of infinity and
analytic arbitrary potential in the neighborhood of the  non-singular endpoint.
We prove the existence of solutions in all these cases, we construct a
 representation of solutions in various series. In addition, in this paper we obtain exact solutions
in closed form for the Friedmann equation with potentials in the form of  constant and
 exponent. These solutions have a different view than the solutions obtained in ref [2-4].

\section{Formulation of the problem. Special solutions. Solutions for the potentials in the form of constant and exponent.}
Let's consider the system of  Friedmann equations for universe filled with a scalar field:

\be
  \begin{cases}\ddot{\varphi}+3H\dot{\varphi}+V'_{\phi}=0,\\
                               H^{2}=\frac{1}{3M^{2}}\ (\frac{\dot{\varphi}^{2}}{2}\ + V(\phi)). \end{cases}\ee

Here $\varphi$ is a scalar field, depending on the time, H is Hubble constant, M and m are real, positive constants, $V(\phi)$ is given continuously differentiable scalar potential. We differentiate with respect to $\varphi$:

\be
  \ 2HH'_{\phi}=\frac{1}{3M^{2}}\ (\dot{\varphi} \frac{d\dot{\varphi}}{d\varphi}\ +V'_{\phi} ).\ee
If $H\equiv0, V<0$, then
 $$
 \int_{\varphi(t_{0})}^{\varphi(t)}\ {\frac{d \varphi}{\sqrt{-2V(\varphi)}}}=\pm(t-t_{0}).$$
If $H\neq0$, then we express from the first equation of (1) H and substitute in (2):
\be
  \dot{\varphi}=-2M^{2}H'_{\phi}.\ee
Finally, by substituting (3) into the second equation of the system (1) and by making replacement: $H=\frac{1}{3}y, \varphi=M\sqrt{\frac{2}{3}}x,U=\frac{3V}{M^{2}}$, we finally obtain:

\be
  \begin{cases}y'_{x}=-\dot{x},\\
                               {y'_{x}}^{2}=y^{2}-U(x). \end{cases}\ee
This system of equations was obtained earlier in [6].
If $U=C^{2}=const$, then a special solution (4) is $y=\pm C, x=C_{1}.$
If $y\neq const$, then:
\be
 \int_{x(t)}^{x(t_{0})}\ {\frac{dx}{y'_{x}}}=t-t_{0}.\ee
Thus, provided that the solution of the second equation (4) is obtained, the first equation can be easily integrated. Let's study the second equation.

1. If U=0, then a solution of (4) is:
$$
y=Ce^{\pm x},  x=\mp ln(Ct+C_{1}).$$

2.If $U=C^{2}, C\neq 0.$ Then the solution is:
$$
y=\pm C ch(x+C_{1}), |x+C_{1}|=arch(cth(\pm Ct+C_{2})).$$

3.If $U=-C^{2}, C\neq 0.$ Then the solution is:
$$
y=\pm C sh(x+C_{1}), x=arsh(tg(\mp Ct+C_{2}))-C_{1}.$$

4.If $U=C_{1}^{2}e^{2C_{2}x}, C_{1}\neq 0.$ Then the solution is:
$$ y=\pm C_{1}e^{C_{2}x}ch u$$
where $u$ is determined by the following equations:

if $|C_{2}|>1$
$$ x=C_{3}-\frac{C_{2}\ln ch(u-arth\frac{1}{C_{2}})+u}{C_{2}^{2}-1},$$

if $C_{2}=\pm 1:$
$$ x=\frac{2u\mp e^{2u}-C_{3}}{4},$$

if $|C_{2}|<1$
$$ x=C_{3}-\frac{C_{2}\ln sh(u-arth C_{2})+u}{C_{2}^{2}-1}.$$

5. $U=-C_{1}^{2}e^{2C_{2}x}, C_{1}\neq 0.$ Then the solution is:
$$ y=\pm C_{1}e^{C_{2}x}sh u$$
where $u$ in the $|C_{2}|>1$ is determined by the following equation:
$$ x=C_{3}-\frac{C_{2}\ln sh(u-arth\frac{1}{C_{2}})+u}{C_{2}^{2}-1}.$$
If $C_{2}=\pm 1:$
$$ x=\frac{2u\pm e^{2u}-C_{3}}{4}.$$
If $|C_{2}|<1$
$$ x=C_{3}-\frac{C_{2}\ln ch(u-arth C_{2})+u}{C_{2}^{2}-1}.$$
In the fourth and fifth cases the dependence y(x) is parametric. We obtain the same dependence for x(t) from (5).

\section{The solution in the form of the series for a quadratic potential in the spiral area (in the neighborhood of x=0).}
Let's consider the potential $V(\varphi)=\frac{m^{2}\varphi^{2}}{2}$. Making the same transformations as in the preparation of (4), and making replacement: $H=\frac{m}{3}y, \varphi=M\sqrt{\frac{2}{3}}x,U=\frac{3V}{M^{2}m^{2}},  t=\frac{t_{1}}{m}$ (hereinafter the subscript 1 is omitted for brevity), we obtain:
$$
  \begin{cases}y'_{x}=-\dot{x},\\
                               {y'_{x}}^{2}=y^{2}-x^{2}. \end{cases}$$
Let's make replacement in this system:
$$y=\frac{2}{u},  x=\frac{2}{u}\cos(\frac{\psi}{2}),  \dot{x}=\frac{2}{u}\sin(\frac{\psi}{2}).$$
For $u=u(\psi)$ we obtain the following equation:
\be
  2u'(u+\sin\psi )=u(\cos\psi -1).\ee
  After receiving the solution of this equation, we will get the parametric dependence y(x) and x(t) with parameter $\psi$. Let's obtain the solution.

\textbf{THEOREM 1.} {\it The solution of equation (6), provided by $\psi\leq\psi_{0}<C-63$ (C is a constant depending on the initial conditions, $\psi_{0}$ is arbitrary constant) is as follows absolutely and uniformly convergent under the same condition series:
\be
  u=\frac{\sin \psi -\psi +C}{2}+\sum_{k=0}^{\infty}a_{k}(\psi),\ee
where
\be a_{0}(\psi)=\int _{-\infty} ^{\psi} \frac{(1-\cos \psi_{1})\sin \psi_{1}d \psi_{1}}{3sin \psi_{1}-\psi_{1}+C},\ee
\be a_{k}(\psi)=-2 \int _{-\infty} ^{\psi}\frac{\sum_{n=0}^{k-1}a_{n}(\psi_{1})a_{k-1-n}'(\psi_{1})d\psi_{1}}{3sin \psi_{1}-\psi_{1}+C},  k \geq 1.\ee

Under the same condition the series consisting of derivatives of this series terms, uniformly and absolutely  converges.}


\begin{proof}
We seek a solution (6) under the conditions:
$$|y|\ll 1 \Rightarrow |x|\ll 1, |u|\gg 1 \Rightarrow |u|\gg \sin \psi.$$
 Under these conditions we neglect $\sin \psi$ compared with $u$ in (6). Then we get the equation:
$$ 2u'=\cos\psi -1,$$
by solving which, we obtain:
$$ u=\frac{\sin \psi -\psi +C}{2}.$$
Therefore, it is advisable to seek the solution of (6) in the form (7). Substituting (7) into (6) we obtain:
\be
  (\cos\psi -1)\sin \psi +\sum_{k=0}^{\infty}a_{k}'(\psi)(3\sin \psi -\psi +C) +2\sum_{k=0}^{\infty}\sum_{n=0}^{k}a_{n}(\psi)a_{k-n}'(\psi)=0.\ee
Equating in (10) the sum of the first summand and zero term of the first series to 0, we get (8). Equating the sum of the k-th term of the first series and the k-1st member of the second series to 0, we obtain (9). We prove that in conditions of the theorem the following estimates are correct:
\be |a'_{k}|\leq \frac{2\cdot 60^{k}}{(k+1)^{2}(C-3-\psi)^{k+1}},\ee
\be |a_{k}|\leq \frac{2\cdot 60^{k}}{(k+1)^{2}(C-3-\psi)^{k}}.\ee
From (8) we get:
$$ |a'_{0}|\leq \frac{2}{C-3-\psi},$$
which corresponds to (11). Produced in (8) integration by parts, we obtain:
$$|a_{0}|\leq |\frac{-\cos \psi+0,25\cos 2\psi}{3\sin \psi -\psi +C}|+|\int _{-\infty} ^{\psi} \frac{(-\cos \psi_{1}+0,25\cos 2\psi_{1})(3\cos \psi_{1}-1)d \psi_{1}}{(3sin \psi_{1}-\psi_{1}+C)^{2}}|\leq $$
$$\leq \frac{25}{4(C-3-\psi)}\leq \frac{25}{256}\leq 2,$$
which corresponds to (12). From (9) we obtain the following chain of inequalities:
$$ |a'_{1}|\leq \frac{8}{(C-3-\psi)^{2}},$$
$$ |a_{1}|\leq \frac{8}{C-3-\psi},$$
$$ |a'_{2}|\leq \frac{64}{(C-3-\psi)^{3}},$$
$$ |a_{2}|\leq \frac{64}{(C-3-\psi)^{2}},$$
which corresponds to (11) and (12). We prove (11) and (12) by induction. Let us assume that these inequalities are correct for $\forall n\leq k-1$. Let us prove it to $n=k$. From (9) for $k\geq 3$ we obtain:
$$ |a'_{k}|\leq \frac{8\cdot 60^{k-1}}{(C-3-\psi)^{k+1}}\sum_{n=0}^{k-1}\frac{1}{(n+1)^2(k-n)^2}.$$
Let's estimate the sum above:
$$\sum_{n=0}^{k-1}\frac{1}{(k-n)^{2}(n+1)^{2}}\leq\frac{2}{k^{2}}+\int_{1}^{k-1}\\\frac{dx}{x^{2}(k-x)^{2}}=$$
 \be =\frac{2}{k^{2}}+\frac{4ln(k-1)}{k^{3}}+\frac{2(k-2)}{k^{2}(k-1}\leq\frac{8}{k^{2}}\leq\frac{15}{(k+1)^{2}}.\ee
 Thus (11) is proved. Integrating (11), we prove (12). For the absolute convergence of the series it is enough that:
$$\lim_{k\rightarrow\infty}\sqrt[k]{|a_{k}|}\leq \frac{60}{C-3-\psi}<1.$$
Consequently,
$$ \psi <C-63,$$
QED. Uniform and absolute convergence of the series, composed of terms which are the derivatives of this series, is proved similarly.
\end{proof}

{\bf Remark.}{\it The solution of (6), provided $\psi \leq \psi_{0}< C-63$ may be represented as follows:}
$$u=\frac{\sin \psi -\psi +C}{2}+\sum_{k=0}^{n}a_{k}(\psi)+o((\frac{60}{C-3-\psi})^{n}),  n\rightarrow\infty.\ $$

Indeed, estimating remaining term of the series with the formula of the sum of a geometric progression, from (12) we get:
$$
|\sum_{k=n+1}^{\infty}a_{k}(\psi)| \leq \frac{|a_{n+1}(\psi)|}{1-\frac{60}{C-3-\psi}} \leq\frac{2\cdot 60^{n+1}}{(n+2)^{2}(C-3-\psi)^{n+1}(1-\frac{60}{C-3-\psi_{0}})}.
$$

\section{The first solution in the form of the series for a quadratic potential in the attractor region.}
\textbf{Theorem 2.} {\it The solution of
\be {y'_{x}}^{2}=y^{2}-x^{2} \ee
under the conditions:
\be 15<x_{2}\leq x\leq x_{-}\leq\sqrt{x_{0}}\leq x_{0}\leq +\infty, x_{0}>3x_{2}^{2},\ee
where
$$ x_{-}=2\sqrt{\frac{x_{0}}{3}}\cos(\frac{1}{3}(\pi-\arccos\frac{3\sqrt{3}}{2\sqrt{x_{0}}}))\geq\sqrt{\frac{x_{0}}{3}}, $$
$x_{2}$ is arbitrary constant, $x_{0}$ is constant depending on the initial data, will be absolutely and uniformly convergent under the same condition series:
\be y=x+\sum_{k=0}^{\infty}a_{k}(x),\ee
where
\be a_{0}(x)=\frac{1}{2}e^{\frac{x^{2}}{2}}\int_{x}^{x_{0}}e^{-\frac{x_{1}^{2}}{2}}dx_{1},\ee
\be
a_{k}(x)=-\frac{1}{2}e^{\frac{x^{2}}{2}}\int_{x}^{x_{0}}e^{-\frac{x_{1}^{2}}{2}}
\sum_{n=0}^{k-1}(a_{n}(x_{1})a_{k-1-n}(x_{1})-a'_{n}(x_{1})a'_{k-1-n}(x_{1}))dx_{1}, k\geq 1.\ee
Under the same condition series, consisting of derivatives of terms of this series, uniformly and absolutely converges.}

\begin{proof}
Substitute (16) in (14):
\be 1+2\sum_{k=0}^{\infty}a'_{k}(x)+\sum_{k=0}^{\infty}\sum_{n=0}^{k}a'_{k-n}(x)a'_{n}(x)=
2x\sum_{k=0}^{\infty}a_{k}(x)+\sum_{k=0}^{\infty}\sum_{n=0}^{k}a_{k-n}(x)a_{n}(x).\ee
We equate the sum of the first summand and zero term of the first series from left to zero member of the first series on the right:
\be a'_{0}(x)=xa_{0}(x)-\frac{1}{2}.\ee
Solving this equation, we get (17). Next equate (19) the sum of
k-th term of the first series on the left and k-1-th member of the second series from left to a similar sum on the right:
\be a'_{k}(x)=xa_{k}(x)+\frac{1}{2}\sum_{n=0}^{k-1}(a_{k-1-n}(x)a_{n}(x)-a'_{k-1-n}(x)a'_{n}(x)), k\geq1.
  \ee
 By solving (21), we obtain (18. Now We estimate $a_{0}$. To do this, consider the following function:
\be f_{1}(x)=\int_{x}^{+\infty}e^{-\frac{x_{1}^{2}}{2}}dx_{1}-\frac{e^{-\frac{x^{2}}{2}}}{x}, x>0.\ee
Apparently
$$ \lim_{x\rightarrow+\infty}f_{1}(x)=0,    f'_{1}(x)=\frac{e^{-\frac{x^{2}}{2}}}{x^{2}}>0.$$
Thus, $f_{1}(x)$, increases and tends to $0$ for $x\rightarrow+\infty$. Therefore,
\be f_{1}(x)<0, x>0.\ee
Now let us consider another function:
\be f_{2}(x)=\int_{x}^{+\infty}e^{-\frac{x_{1}^{2}}{2}}dx_{1}-\frac{2e^{-\frac{x^{2}}{2}}}{x+\sqrt{x^{2}+4}}, x>0.\ee
Apparently
$$ \lim_{x\rightarrow+\infty}f_{2}(x)=0,    f'_{2}(x)=\frac{e^{-\frac{x^{2}}{2}}(x\sqrt{x^{2}+4}-x^{2}-2)}{(x+\sqrt{x^{2}+4})\sqrt{x^{2}+4}}<0.$$
Thus, $f_{2}(x)$, decreases and tends to $0$ for $x\rightarrow+\infty$. Therefore,
\be f_{2}(x)>0, x>0.\ee
From (22) - (25) it follows that:
\be  \frac{1}{x+\sqrt{x^{2}+4}}-\frac{1}{2x_{0}}\leq a_{0}(x)=|a_{0}(x)|\leq \frac{1}{2x},  x>0.\ee
Now we estimate the $a'_{0}(x)$. From (20) and (26) you can see that $a'_{0}(x)\leq 0$. Therefore,
\be |a'_{0}(x)|=\frac{1}{2}-xa_{0}(x)\leq \frac{1}{2}-\frac{1}{1+\sqrt{1+\frac{4}{x^{2}}}}+\frac{2x}{x_{0}}\leq \frac{1}{2}(\frac{1}{x^{2}+1}+\frac{x}{x_{0}})\leq \frac{1}{2}(\frac{1}{x^{2}}+\frac{x}{x_{0}}).\ee
Impose the following limit on $a'_{0}(x)$:
\be |a'_{0}(x)|\leq \frac{1}{2x},  x>0.\ee
This is possible if
\be \frac{1}{x^{2}}+\frac{x}{x_{0}}\leq \frac{1}{x}\ee
Multiplying both sides of (29) to $x^{2}x_{0}$, we obtain a cubic inequality:
\be x^{3}-xx_{0}+x_{0}\leq0.\ee
Solving (30) according to the formula Cardano, we obtain for $x>0$:
\be x_{+}\leq x\leq x_{-},\ee
where
\be x_{\pm}=2\sqrt{\frac{x_{0}}{3}}\cos(\frac{1}{3}(\pi\pm \arccos\frac{3\sqrt{3}}{2\sqrt{x_{0}}})),  x_{0}\geq 6,75.\ee
By analyzing the obtained solution, we get:
$$ \sqrt{\frac{x_{0}}{3}}\leq x_{-}\leq \sqrt{x_{0}},$$
\be 1\leq x_{+}\leq 1,5\leq x_{-}\leq +\infty.\ee
Let's introduce the notation:
$$ b_{k}=\max \{|a_{k}|, |a'_{k}|\}.$$
We prove that for $\forall k\geq 0, \forall x,$ satisfying (31)
\be b_{k}\leq\frac{15^{k}}{2x^{k+1}(k+1)^{2}}.\ee
For $k=0$ it is already proved. From (18) and (23) we obtain:
\be |a_{k}|\leq \frac{1}{x}\sum_{n=0}^{k-1}b_{k-1-n}(x)b_{n}(x), k\geq1.\ee
From (21):
\be |a'_{k}|\leq 2\sum_{n=0}^{k-1}b_{k-1-n}(x)b_{n}(x), k\geq1.\ee
Considering (35), (36), (31) and (33), we get:
\be b_{k}\leq 2\sum_{n=0}^{k-1}b_{k-1-n}(x)b_{n}(x), k\geq1.\ee
From (26), (28) and (37):
$$ b_{1}\leq \frac{1}{2x^{2}},  b_{2}\leq \frac{1}{x^{3}},$$
that satisfies (34). Now we prove (34) by induction method. Let's suppose that (34) holds for $\forall n\leq k-1$. Now we now prove that (34) is correct for $n=k$. From (37), (34) and (13) for $\forall k\geq 3:$
\be b_{k}\leq\frac{15^{k-1}}{2x^{k+1}}\sum_{n=0}^{k-1}\frac{1}{(k-n)^{2}(n+1)^{2}}\leq \frac{15^{k-1}\cdot 8}{2x^{k+1}k^{2}}\leq \frac{15^{k}}{2x^{k+1}(k+1)^{2}}.\ee
Thus (34) is proved. Further, for the convergence of the series (16) and its formal derivative it is sufficient:
$$ \lim_{k\rightarrow\infty}\sqrt[k]{b_{k}}\leq \frac{15}{x}<1.$$
So, $x>15$. In addition, from (15) and (33) you can see that the inequality $x_{0}>3x_{2}^{2}.$ is sufficient.
\end{proof}

{\bf Note 1.}{\it The solution of (14) under the condition (15) can be represented as follows:}
$$y=x+\sum_{k=0}^{n}a_{k}(x)+o((\frac{15}{x})^{n}),  n\rightarrow\infty.\ $$
Indeed, assessing the remaining term of the series with the formula of geometric progression sum, from (34) we get:
$$
|\sum_{k=n+1}^{\infty}a_{k}(x)| \leq \frac{|a_{n+1}(x)|}{1-\frac{15}{x}} \leq\frac{16\cdot 15^{n+1}}{2(n+2)^{2}x^{n+2}}.
$$

{\bf  Note 2.}{\it Because the solution of equation (14) is even there is a similar solution near the line $y=-x$, and also a solution obtained by replacing $y$ with $-y$.}

{\bf Note 3.}{\it If in (15) we put $x_{0}=+\infty$, then all the arguments remain valid, and we get a "clean" attractor solution which exists in a neighborhood of infinity and has a line $y=x$ as its asymptote. The attempt to find a solution in the form of a series
$$ y=x+ \sum _{k=0}^{\infty}\frac{a_{k}}{x^{k}}$$
fails - series is divergent.}

\section{The second solution in the form of the series for a quadratic potential in the attractor region.}
\textbf{Theorem 3.} {\it The solution of Cauchy problem
\be {y'_{x}}^{2}=y^{2}-x^{2},  y(x_{0})=\sqrt{x_{0}^{2}+1} \ee

under conditions:
\be 4\leq x\leq x_{0}\leq +\infty,  x_{0}>4 \ee
 is the next absolutely and uniformly convergent under the same condition series:
\be y=\sqrt{x^{2}+1}+\sum_{k=0}^{\infty}a_{k}(x),\ee
where
\be a_{0}(x)=-xe^{\frac{x^{2}}{2}}\int_{x}^{x_{0}}\frac{e^{-\frac{x_{1}^{2}}{2}}dx_{1}}{2x_{1}^{2}\sqrt{x_{1}^{2}+1}},\ee
\be
a_{k}(x)=-xe^{\frac{x^{2}}{2}}\int_{x}^{x_{0}}\frac{e^{-\frac{x_{1}^{2}}{2}}\sqrt{x_{1}^{2}+1}
\sum_{n=0}^{k-1}(a_{n}(x_{1})a_{k-1-n}(x_{1})-a'_{n}(x_{1})a'_{k-1-n}(x_{1}))dx_{1}}{2x_{1}^{2}}, k\geq 1.\ee
Under the same condition series, consisting of these series terms derivatives, uniformly and absolutely converge.}

\begin{proof}
Substituting (41) into (39):
$$ \frac{2x}{\sqrt{x^{2}+1}}\sum_{k=0}^{\infty}a'_{k}(x)+\sum_{k=0}^{\infty}\sum_{n=0}^{k}a'_{k-n}(x)a'_{n}(x)=
\frac{1}{x^{2}+1}+2\sqrt{x^{2}+1}\sum_{k=0}^{\infty}a_{k}(x)+$$

\be+\sum_{k=0}^{\infty}\sum_{n=0}^{k}a_{k-n}(x)a_{n}(x).\ee

We equate the zero term of the first series to the left to the sum of the first summand and zero term of the first series on the right:
\be a'_{0}(x)=(x+\frac{1}{x})a_{0}(x)+\frac{1}{2x^{2}\sqrt{x^{2}+1}}.\ee
Solving this equation, we get (42). Next we equate in (44) the sum of
k-th term of the first series on the left and k-1-th term of the second series from left to a similar sum on the right:
\be a'_{k}(x)=(x+\frac{1}{x})a_{k}(x)+\frac{\sqrt{x^{2}+1}}{2x}\sum_{n=0}^{k-1}(a_{k-1-n}(x)a_{n}(x)-a'_{k-1-n}(x)a'_{n}(x)), k\geq1.
  \ee
Solving (46), we obtain (43). Now we estimate $a_{0}$ and $a'_{0}$. From (22), (23) and (42):
\be -\frac{1}{2x^{3}}\leq-\frac{1}{2x^{2}\sqrt{x^{2}+1}}\leq a_{0}\leq 0.\ee
From (45) and (47):
\be -\frac{1}{2x^{3}\sqrt{x^{2}+1}}\leq a'_{0}\leq \frac{1}{2x\sqrt{x^{2}+1}}\leq \frac{1}{2x^{2}}.\ee
We introduce the notation:
$$ b_{k}=\max \{|a_{k}|, |a'_{k}|\}.$$
Then from (40), (47) and (48):
\be b_{0}\leq \frac{1}{2x^{2}}.\ee
Further, from (43) for $\forall k\geq1$:
$$ |a_{k}|\leq \frac{\sqrt{17}}{4x}\sum_{n=0}^{k-1}b_{n}b_{k-1-n}.$$
From (46) we get:
$$ |a'_{k}|\leq \frac{33\sqrt{17}}{64}\sum_{n=0}^{k-1}b_{n}b_{k-1-n}.$$
Consequently
\be b_{k}\leq \frac{33\sqrt{17}}{64}\sum_{n=0}^{k-1}b_{n}b_{k-1-n}.\ee
We prove that for $\forall k\geq 0, \forall x,$ satisfying (40)
\be b_{k}\leq\frac{15,95^{k}}{2x^{2k+2}(k+1)^{2}}.\ee
For $k=0$ it is already proved. From (49) and (50) we get:
$$ b_{1}\leq \frac{33\sqrt{17}}{256x^{4}},  b_{2}\leq \frac{18513}{16384x^{6}},$$
that satisfies (51). Let's prove (34) by induction. We suppose that (51) is true for $\forall n\leq k-1$. Now we prove that (51) for $n=k$.  From (50), (51) and (13) for $\forall k\geq 3:$
\be b_{k}\leq\frac{33\sqrt{17}\cdot15,95^{k-1}\cdot8}{256x^{2k+2}k^{2}}\leq \frac{495\sqrt{17}\cdot15,95^{k-1}}{256x^{2k+2}(k+1)^{2}}\leq \frac{15,95^{k}}{2x^{2k+2}(k+1)^{2}},\ee
This proves (51). Further, for the convergence of the series (41) and its formal derivative it is sufficient:
$$ \lim_{k\rightarrow\infty}\sqrt[k]{b_{k}}\leq \frac{15,95}{x^{2}}<1.$$
Thus, $x\geq4$ is enough.
\end{proof}

{\bf Note 1.}{\it Solution of the problem (39) under the condition (40) can be represented as follows:}
$$y=\sqrt{x^{2}+1}+\sum_{k=0}^{n}a_{k}(x)+o((\frac{15,95}{x^{2}})^{n}),  n\rightarrow\infty.\ $$
Indeed, assessing the remaining term of the series with the formula of geometric progression sum, from (51) we get:
$$
|\sum_{k=n+1}^{\infty}a_{k}(x)| \leq \frac{|a_{n+1}(x)|}{1-\frac{15,95}{x^{2}}} \leq\frac{15,95^{n+1}}{2(n+2)^{2}x^{2n+4}(1-\frac{15,95}{x^{2}})}\leq \frac{320\cdot 15,95^{n+1}}{2(n+2)^{2}x^{2n+4}}.
$$
Notes (2) and (3) of Theorem 2 also remain valid.

\section{The solution in the form of a series in the neighborhood of $x=\infty$ for any continuous potential, growing no faster than exponential.}
\textbf{DEFINITION.} {\it Function $y=f(x)$ will be called a slow-growing, if
$$\exists \alpha <1     \exists x_{0}\exists C_{1}>0 \forall x\geq x_{0} $$

$$|f(x)|\leq C_{1}^{2}e^{2\alpha x}.$$
}

\textbf{THEOREM 4.} {\it The solution of equation
\be (y'_{x})^{2}=y^{2}-U(x),\ee
where $U(x)$ is given continuous slowly increasing function, under condition:
\be x\geq x_{1}=max\{\frac{1}{1-\alpha}\ln\frac{C_{1}\sqrt{2}(3-2\alpha)}{C(1-\alpha)};x_{0}\}\ee
will be the next series, which are absolutely and uniformly convergent together with its derivative :
\be y=Ce^{x}+\sum_{k=0}^{\infty}a_{k}(x)
  ,\ee
  where $C$ is an arbitrary, positive, depending on the initial constant,
  \be  a_{0}(x)=\frac{e^{x}}{2C}\int_{x}^{+\infty}U(x_{1})e^{-2x_{1}}dx_{1},\ee

\be a_{k}(x)=-\frac{e^{x}}{2C}\int_{x}^\infty\\\sum_{m=0}^{k-1}(a_{k-1-m}(x_{1})a_{m}(x_{1})-a'_{k-1-m}(x_{1})a'_{m}(x_{1}))e^{-2x_{1}}dx_{1}, k\geq1.
  \ee}

\begin{proof}
Substituting the series (55) into equation (53):
\be C^{2}e^{2x}+2Ce^{x}\sum_{k=0}^{\infty}a'_{k}(x)+\sum_{k=0}^{\infty}\sum_{m=0}^{k}a'_{k-m}a'_{m}=
C^{2}e^{2x}+2Ce^{x}\sum_{k=0}^{\infty}a_{k}(x)+\sum_{k=0}^{\infty}\sum_{m=0}^{k}a_{k-m}a_{m}-U(x).\ee

Equating the zero term of the first series on the left to the sum of the zero term of the first series on the right and the last
 summand on the right:
\be a'_{0}=a_{0}-\frac{U(x)e^{-x}}{2C}. \ee
Solvinq (59) and equating the integration constant to zero, we obtain (56).
Further equate in (58) the sum of
k-th term of the first series on the left and k-1-th term of the second series from left to the similar sum on the right:
\be a'_{k}(x)=a_{k}(x)+\frac{1}{2C}\sum_{m=0}^{k-1}(a_{k-1-m}(x)a_{m}(x)-a'_{k-1-m}(x)a'_{m}(x))e^{-x}, k\geq1.
  \ee
Solvinq (60), again assuming constant of integration equal to zero, we obtain (57).
 Now we prove the convergence of the resulting series and series consisting of derivatives of terms of this series. From (56) for $x\geq x_{0}$, we get:
\be |a_{0}|\leq \frac{C_{1}^{2}e^{(2\alpha -1)x}}{4C(1-\alpha)}.\ee
From (59):
\be |a'_{0}|\leq \frac{C_{1}^{2}e^{(2\alpha -1)x}}{2C}(1+\frac{1}{2(1-\alpha)}).\ee
We introduce the notation:
$$ max\{|a_{k}|,|a'_{k}|\}=b_{k}.$$
From (61) and (62) it follows:
\be b_{0}\leq \frac{C_{1}^{2}e^{(2\alpha -1)x}}{2C}(1+\frac{1}{2(1-\alpha)}).\ee
From (57) and (60) for $x\geq x_{0}$:
\be b_{1}\leq \frac{C_{1}^{4}e^{(4\alpha -3)x}}{4C^{3}}(1+\frac{1}{2(1-\alpha)})^{2}(1+\frac{1}{4(1-\alpha)})\leq \frac{C_{1}^{4}e^{(4\alpha -3)x}}{4C^{3}}(1+\frac{1}{2(1-\alpha)})^{3},\ee
\be b_{2}\leq \frac{C_{1}^{6}e^{(6\alpha -5)x}}{4C^{5}}(1+\frac{1}{2(1-\alpha)})^{4}(1+\frac{1}{6(1-\alpha)})\leq \frac{C_{1}^{4}e^{(4\alpha -3)x}}{4C^{3}}(1+\frac{1}{2(1-\alpha)})^{5}.\ee
We make the induction hypothesis: suppose that for any m<k
\be
b_{m}\leq \frac{C_{1}^{2m+2}e^{(2(\alpha -1)(m+1)+1)x}15^{m}}{2^{m+1}C^{2m+1}(m+1)^{2}}(1+\frac{1}{2(1-\alpha)})^{2m+1}.
\ee
Obviously, что $b_{0}, b_{1}, b_{2}$ satisfy (66). Using the hypothesis (66) from (57) and (60) we obtain:
 \be
 b_{k}\leq \frac{C_{1}^{2k+2}e^{(2(\alpha -1)(k+1)+1)x}15^{k-1}}{2^{k+1}C^{2k+1}}(1+\frac{1}{2(1-\alpha)})^{2k+1}\sum_{m=0}^{k-1}\frac{1}{(k-m)^{2}(m+1)^{2}}.
 \ee
 Using (13), we prove (66) for $m=k$.
  For the convergence of the series it is enough:
 $$\lim_{k\rightarrow\infty}\sqrt[k]{b_{k}}\leq \frac{15C_{1}^{2}e^{2(\alpha -1)x}}{2C^{2}}(1+\frac{1}{2(1-\alpha)})^{2}<1.
  $$
  This proves the sufficiency of (54).
\end{proof}

{\bf Note 1.}{\it The solution of (53) under the condition (54) can be represented as follows:
\be y=Ce^{x}+\sum_{k=0}^{n}a_{k}(x)+o((\frac{15C_{1}^{2}e^{2(\alpha -1)x}}{2C^{2}}(1+\frac{1}{2(1-\alpha)})^{2})^{n}),  n\rightarrow\infty.\ \ee}
Indeed, assessing the remaining term of the series with the sum of the geometric progression formula, we get:
$$
|\sum_{k=n+1}^{\infty}a_{k}(x)| \leq |a_{n+1}(x)|(1-\frac{15C_{1}^{2}e^{2(\alpha -1)x}}{2C^{2}}(1+\frac{1}{2(1-\alpha)})^{2})^{-1},
$$
This proves (68).

{\bf Note 2.}{\it It is possible to build a similar solution in the neighborhood of $x=-\infty$, the solution obtained by replacing $y$ with $-y$ is also true.}

\section{The solution in the form of a Taylor series in the neighborhood of $x=x_{0}$ for any analytical potentials.}
Let's pose the Cauchy problem:
\be {y'_{x}}^{2}=y^{2}-U(x),   y(x_{0})=y_{0},  y_{0}^{2}>U(x_{0}),\ee
where $U(x)$ is defined function that is analytic in a neighborhood of $x=x_{0}$ and having a positive radius of convergence $R$:
\be U(x)=\sum_{m=0}^{\infty}b_{m}(x-x_{0})^{m}.\ee
\textbf{THEOREM 5.} {\it Solution of problem (69) are the series:
\be y=\sum_{m=0}^{\infty}a_{m}(x-x_{0})^{m},\ee
where
$$ a_{0}=y_{0},  a_{1}^{2}=y_{0}^{2}-b_{0},$$
\be a_{m+1}=\frac{1}{2a_{1}(m+1)}(\sum_{k=0}^{m}a_{k}a_{m-k}-\sum_{k=1}^{m-1}(k+1)(m-k+1)a_{k+1}a_{m-k+1}-b_{m}),
m\geq 1.\ee
For $m=1$ we consider that the second sum in the parenthesis in (72) equal zero. The radius of convergence (71) is estimated as follows:
\be R_{1}\geq\frac{2|a_{1}|}{S},\ee
$$ S= \max (64a_{2}^{2}; 9(2|a_{1}a_{3}|)^{\frac{2}{3}};  \frac{4|a_{1}|}{R};$$
\be (\frac{16|y_{0}|}{81}+\frac{19a_{1}^{2}}{324}+6+M)^{2}),\ee
\be M=\max (\frac{b_{m}R^{m+1}(m+1)^{2}}{2^{m}|a_{1}|}),  m\geq3.\ee
}

  \begin{proof}
  Substituting (71) and (70) to (69) and equating the coefficients of equal powers of $x-x_{0}$ we get (72). To estimate the radius of convergence of the series (71) we will state the induction hypothesis. Suppose that $\forall 2\leq k\leq m$
\be |a_{k}|\leq\frac{S^{k-1,5}}{k^{3}(2|a_{1}|)^{k-2}},\ee
where $S$ is positive constant to be determined. We prove this statement for $k=m+1$ (in addition, this automatically implies the validity of (73)). From (76) for $k=2,3$ we have:
\be S\geq 64a_{2}^{2},\ee
\be S\geq 9(2|a_{1}a_{3}|)^{\frac{2}{3}}.\ee
From (72), to prove (76) for $ k = m + 1 $ it is enough:
 $$|a_{m+1}|\leq \frac{1}{2|a_{1}|(m+1)}(\frac{2|y_{0}|S^{m-1,5}}{m^{3}(2|a_{1}|)^{m-2}}+
 \frac{S^{m-2,5}}{(m-1)^{3}(2|a_{1}|)^{m-2}}+
\sum_{k=2}^{m-2}\frac{S^{m-3}}{k^{3}(2|a_{1}|)^{m-4}(m-k)^{3}}+$$
\be +\sum_{k=1}^{m-1}\frac{S^{m-1}}{(k+1)^{2}(2|a_{1}|)^{m-2}(m-k+1)^{2}}+|b_{m}|)
\leq \frac{S^{m-0,5}}{(m+1)^{3}(2|a_{1}|)^{m-1}}, m\geq 3.\ee
From (79), in which we believe that the first sum for $m=3$ is equal to zero, we get one more condition for $S$:
$$ S\geq (\frac{2|y_{0}|(m+1)^{2}}{m^{3}\sqrt{S}}+
 \frac{(2a_{1})^{2}(m+1)^{2}}{(m-1)^{3}S\sqrt{S}}+
\frac{(2a_{1})^{2}(m+1)^{2}}{S^{2}}\sum_{k=2}^{m-2}\frac{1}{k^{3}(m-k)^{3}}+$$
\be +(m+1)^{2}\sum_{k=1}^{m-1}\frac{1}{(k+1)^{2}(m-k+1)^{2}}+\frac{|b_{m}|(2|a_{1}|)^{m-2}(m+1)^{2}}{S^{m-1}})^{2}, m\geq3.\ee
We estimate the first sum of (80) with $m\geq 4$:
 $$(m+1)^{2}\sum_{k=2}^{m-2}\frac{1}{k^{3}(m-k)^{3}}\leq (m+1)^{2}\int_{2}^{m-1}\frac{dx}{(x-1)^{3}(m-x)^{3}}=\frac{(m+1)^{2}}{(m-1)^{3}}(1-\frac{1}{(m-2)^{2}})+$$
$$+\frac{6(m+1)^{2}}{(m-1)^{4}}(1-\frac{1}{m-2})+\frac{12(m+1)^{2}\ln(m-2)}{(m-1)^{5}}\leq
\frac{(m+1)^{2}}{(m-1)^{3}}+\frac{18(m+1)^{2}}{(m-1)^{4}}=$$
\be =\frac{(m+1)^{2}(m+17)}{(m-1)^{4}}\leq (1+\frac{18}{m-1})(1+\frac{2}{m-1})^{2}\frac{1}{m-1}\leq\frac{175}{27}\leq 7.\ee
For the second sum we get:
$$ (m+1)^{2}\sum_{k=1}^{m-1}\frac{1}{(k+1)^{2}(m-k+1)^{2}}\leq(m+1)^{2}\int_{1}^{m}\frac{dx}{x^{2}(m+1-x)^{2}}=$$
\be=\frac{4\ln m}{m+1}+\frac{2(m-1)}{m}\leq 6.\ee
In (80) by substituting this sum with 6, we will strengthen the inequality. Then from this inequality it will follows: $S\geq36$. Last term from (80) presents from itself a sequence, for convergence of which to the zero it is sufficient that
$$ S>\frac{2|a_{1}|}{R}.$$

Suppose that in this sequence
\be S=\frac{4|a_{1}|}{R}.\ee
Then the sequence will be limited. Let $M$ to be its maximum. Let's replace the last term on the right side (80) with $M$. Also we replace $S$ with $36$ in the first three terms and $m$ with $3$ in the first two terms. Further, using the estimates (81) and (82),
we only strengthen the inequality:

\be S\geq (\frac{16|y_{0}|}{81}+\frac{19a_{1}^{2}}{324}+6+M)^{2}.\ee
Using (77), (78), (83) and (84) we prove (74).
\end{proof}

\section*{Acknowledgments}
This work was begun on the Steklov Mathematical Institute seminar. We are thankfull to all
the participants for active discussion of the work. Our special thanks to  Igor Vasilyevich Volovich for the problem statement and the number of valuable advices.


\end{document}